\documentclass[prb,twocolumn,amssymb,showpacs,superscriptaddress]{revtex4}
\usepackage{graphicx,amsmath}
\usepackage{color}
\begin{document}
\title{Sum-rules and bath-parametrization for quantum cluster theories}
\author{Erik Koch}
%\email{E.Koch@fz-juelich.de}
\affiliation{Institut f\"ur Festk\"orperforschung,
             Forschungszentrum J\"ulich, 52425 J\"ulich, Germany}
\author{Giorgio Sangiovanni}
\author{Olle Gunnarsson}
\affiliation{Max-Planck-Institut f\"ur Festk\"orperforschung, 70569 Stuttgart, Germany}
\date{\today}
\begin{abstract}
We analyze cellular dynamical mean-field theory (CDMFT) and the dynamical cluster approximation (DCA). We derive exact sum-rules for the hybridization functions and give examples for DMFT, CDMFT, and DCA. For impurity solvers based on a Hamiltonian, these sum-rules can be used to monitor convergence of the bath-parametrization. We further discuss how the symmetry of the cluster naturally leads to a decomposition of the bath Green matrix into irreducible components, which can be parametrized independently, and give an explicit recipe for finding the optimal bath-parametrization. 
As a benchmark we revisit the one-dimensional Hubbard model. We carefully analyze the evolution of the density as a function of chemical potential and find that, close to the Mott transition, convergence with cluster size is unexpectedly slow.
In two dimensions we find, that we need so many bath-sites to obtain a reliable parametrization that Lanczos calculations are hardly feasible with current computers. For such large baths our symmetry-adapted approach should prove crucial for finding a reliable bath-parametrization.
\end{abstract}
% 71.10.-w	Theories and models of many-electron systems
% 71.10.Fd  Lattice fermion models (Hubbard model, etc.)
% 71.15.-m	Methods of electronic structure calculations 
% 71.27.+a	Strongly correlated electron systems; heavy fermions
% 71.30.+h  Metal-insulator transitions and other electronic transitions
\maketitle

\newcommand{\Gbath}{\boldsymbol{\cal G}}\newcommand{\Gb}{\mathcal{G}^{-1}}
\newcommand{\Gclust}{\mathbf{G}_c}
\newcommand{\Glatt}{\mathbf{G}}
\newcommand{\rsuper}{\tilde{\mathbf{r}}}
\newcommand{\ksmall}{\tilde{\mathbf{k}}}
\newcommand{\qsmall}{\tilde{\mathbf{q}}}

\section{Introduction}

Strongly correlated materials are characterized by the interplay of kinetic energy and sizable short-range electronic repulsion, which cannot be described with single-particle approaches or standard perturbative theories.
The Dynamical Mean Field Theory\cite{DMFTpapers,DMFTreview} (DMFT) has proven extremely powerful, in particular for the description of the correlation-driven Mott transition. Yet DMFT is strictly local. 
A number of recent developments are aimed at overcoming this limitation and are greatly contributing to the understanding of the physics of strongly correlated systems.\cite{DCA,sasha00,CDMFT,Potthoff,review05,DGammaA,dualfermion}
In quantum cluster theories\cite{review05} the $k$-dependence is introduced by considering a small number of sites, instead of the single correlated site of DMFT, and embedding them in a bath, i.e., a dynamical mean-field host determined self-consistently. This is a good approximation when the self-energy is reasonably localized, which can be systematically improved by considering larger and larger clusters.\cite{maierPRL95}

Cluster extensions of DMFT are not unique. Here we consider the two main flavors, cellular dynamical mean-field theory\cite{CDMFT} (CDMFT) and the dynamical cluster approximation\cite{DCA} (DCA).
Finite temperature Monte Carlo provides an efficient cluster solver, however, to obtain spectra on the real-frequency axis, data for imaginary time must be analytically continued, usually applying maximum entropy. If the spectrum has structures on a small energy scale, this approach can lead to problems. It is then natural to explore Hamiltonian-based solvers, like the Lanczos method, which directly give result on the real-frequency axis. 

Lanczos is extensively applied to DMFT and results are reliable and very accurate.
The critical step in such calculations is the fitting of the bath degrees-of-freedom. In cluster methods this step gets more involved. This is one of the reasons why it found few application for cluster methods: To the best of our knowledge, it has so far only been used for CDMFT calculations of very small clusters.
Our aim is therefore to give a systematic formulation, exploiting symmetries to arrive at an optimal parametrization of the bath. In addition we derive exact sum-rules for the bath Green's functions.
As an application, we investigate the one-dimensional Hubbard model with CDMFT in detail,\cite{bolech,esperti,civelliPhD} paying particular attention to the convergence with the size of both bath and cluster.  

The paper is organized as follows: To fix the notation, in section \ref{SecMethod}, we give a unified formulation of CDMFT and DCA in a formalism using a Hamiltonian solver. In section \ref{SecSumrule} we derive general sum-rules for the hybridizations and give examples for DMFT, CDMFT, and DCA. In section \ref{SecSymmetry} we discuss how the symmetry of the cluster naturally leads to a decomposition of the bath Green matrix into irreducible components, which can be parametrized independently. We give an explicit recipe for finding the optimal bath-parametrization and discuss how this approach relates to the technique of cluster replica. In section \ref{SecEsperti} we analyze how the cluster approaches work for the Hubbard chain and carefully reexamine the evolution of the density as a function of chemical potential. In section \ref{SecConclusions} we give our conclusions and an outlook.

\section{Method and notation}\label{SecMethod}
We consider the Hubbard model
% for extended interaction often mean-field terms are put into Hc
% cf. Bolech eqn above (7)
% for our sum-rule to hold Hc and H(k) must be consistent!
% see also comments at Hcluster
\begin{equation}\label{H}
 H=-\sum_{ij\sigma} 
     t_{ij}\,c^\dagger_{i\sigma}c^{\phantom{\dagger}}_{j\sigma}
    + U\sum_i n_{i\uparrow}n_{i\downarrow}\,.
  % + V\sum_{\rangle ij\langle} n_i n_j
\end{equation}
To fix the notation we briefly sketch the self-consistency loop for cellular DMFT (CDMFT) and the dynamical cluster approximation (DCA) using, e.g., exact diagonalization as impurity solver. Let $N_c$ be the number of cluster-sites, $N_b$ the number of bath-sites. For simplicity we suppress spin-indices.

Given an $N_c\times N_c$ bath Green matrix $\Gbath^{-1}$,
\begin{enumerate}
\item 
 fit parameters of an Anderson model with $N_b$ bath-sites  
 \begin{equation}\label{bathfitting}
  \Gbath^{-1}_\mathrm{And}(\omega)\approx
    \omega+\mu-\mathbf{H}_c
     -\mathbf{\Gamma}\,[\omega-\mathbf{E}]^{-1}\mathbf{\Gamma}^\dagger
 \end{equation}
%which is usually done on the imaginary axis introducing a fictitious temperature.
to $\Gbath^{-1}$, where $\mathbf{\Gamma}$ is the $N_c\times N_b$-dimensional hybridization matrix, and $\mathbf{E}$ the $N_b\times N_b$-dimensional bath-matrix. $\mathbf{H}_c$ is specified below,
\item\label{solve}
 solve the $N_c+N_b$-site Anderson model $H_\mathrm{And}$ (specified below) to obtain the $N_c\times N_c$ cluster Green matrix $\mathbf{G_c}$,
\item
 get the cluster self-energy matrix 
 \begin{equation}
  \boldsymbol{\Sigma}_c(\omega)=\Gbath^{-1}(\omega)-\Gclust^{-1}(\omega)\,,
 \end{equation}
\item
 calculate the local Green matrix for the cluster by integrating over the reduced Brillouin-zone of the cluster
 \begin{equation}
  \Glatt(\omega)=
   \int d\ksmall\; 
    \Big(\omega+\mu-\mathbf{H}(\ksmall)-\boldsymbol{\Sigma}_c(\omega)\Big)^{-1}\,,
 \end{equation}
where $\mathbf{H}(\ksmall)$ is the single-electron part of the of the Hubbard Hamiltonian (\ref{H}) in the reduced Brillouin-zone of the cluster, 
%It is only defined up to local phase factors on the cluster-sites. The choice of these gauge phases characterizes the cluster approximation, e.g., CDMFT or DCA.
% % eqn.10 in Maier00
\item
 determine the new bath Green matrix (self-consistency condition)
 \begin{equation}
  \Gbath^{-1}(\omega)=\mathbf{\Sigma}_c(\omega)+\Glatt^{-1}(\omega)\,.
 \end{equation}
\end{enumerate}
These steps are iterated to self-consistency.

The Anderson model to be solved in step \ref{solve} is given by
\begin{equation}\label{Hand}
  H_\mathrm{And}=H_\mathrm{clu}
   +\sum_{lm,\sigma} E_{lm,\sigma}\,a^\dagger_{l\sigma}a^{\phantom{\dagger}}_{m\sigma}
   +\sum_{li,\sigma} \Gamma_{il} \left(a^\dagger_{l\sigma}c_{i\sigma} + \mathrm{H.c.}\right)
\end{equation}
where the operator $a^\dagger_{l\sigma}$ creates an electron of spin $\sigma$ on bath-site $l$.
The cluster Hamiltonian $H_\mathrm{clu}$ is obtained from the original Hamiltonian (\ref{H}) by transforming to the reciprocal space of the super-lattice of clusters, and projecting to the cluster. Writing the single-electron part of $H(\ksmall)$ as the matrix $\mathbf{H}(\ksmall)$, the single-electron part of $H_\mathrm{clu}$ is given by
\begin{equation}\label{Hc}
 \mathbf{H}_c = \int d\ksmall\,\mathbf{H}(\ksmall)\,.
\end{equation}
The interaction terms are simply those of (\ref{H}), restricted to the cluster.
% Longer-range interactions $\mathbf{U}^{\sigma\sigma'}_{ij}(\qsmall)$ in $H$ become $\mathbf{U}^{\sigma\sigma'}_{ij}=\int d\qsmall\,\mathbf{U}^{\sigma\sigma'}_{ij}(\qsmall)$ on the cluster.

The Hamiltonian $H(\ksmall)$ in the reciprocal space of the super-lattice $\{\rsuper\}$ of clusters can be obtained by changing to the basis of operators 
\begin{equation}\label{FTCDMFT}
 \tilde{c}^\mathrm{CDMFT}_{\mathbf{R}_i\sigma}(\ksmall)
 =\sum_{\rsuper} e^{-i\ksmall\rsuper}\,c_{\rsuper+\mathbf{R}_i,\sigma}\,.
\end{equation} 
The resulting quantum cluster approximation is CDMFT.
Alternatively, we can start from the operators in the reciprocal space of the {\em lattice} to obtain
\begin{equation}\label{FTDCA}
 \tilde{c}^\mathrm{DCA}_{\mathbf{R}_i\sigma}(\ksmall)
 =\sum_{\rsuper} 
   e^{-i\ksmall(\rsuper+\mathbf{R}_i)}\,c_{\rsuper+\mathbf{R}_i,\sigma}\,.
\end{equation}
Now we obtain the DCA. The choice of the operators in the two approaches differs just by local phase factors. In CDMFT this gauge\cite{kohn64} is chosen such that phases appear only in matrix elements involving different clusters. Thus all matrix elements on the cluster are the same as in the original Hamiltonian. The price for retaining the original matrix elements on the cluster is a breaking of the translation-symmetry of the original lattice.
DCA opts instead to retain this symmetry by distributing the phase change uniformly over the cluster-sites. The price for retaining translation-invariance is that the matrix elements in the cluster Hamiltonian differ from those in the original Hamiltonian (coarse graining).
In both cases, CDMFT and DCA, the eigenvalues of $\mathbf{H}(\ksmall)$ are identical to the eigenvalues of the non-interacting part of $H$

\section{Hybridization sum-rules}\label{SecSumrule}
While the most general parametrization for the bath is given by expression (\ref{bathfitting}),\cite{sasha00,bolech} we can always diagonalize the hopping matrix $\mathbf{E}$ among the bath-sites to obtain
\begin{equation}\label{diagbath}
 \Gbath_\mathrm{And}^{-1}(\{\varepsilon_l,\mathbf{V}_l\}; \omega) =
   \omega+\mu-\mathbf{H}_c-\sum_l\frac{\mathbf{V}_l\,\mathbf{V}_l^\dagger}{\omega-\varepsilon_l}\,.
\end{equation}
The hybridization matrix is then given by the tensor product of the vectors $\mathbf{V}_l$, where % cf. cbath.f
\begin{equation}
 V_{l,i}=\sum_m \Gamma_{i,m}\,\phi_{l,m}
\end{equation}
and $\mathbf{\phi}_l$ are the eigenvectors of $\mathbf{E}$ with eigenvalues $\varepsilon_l$.

To obtain sum-rules for the hybridizations, we write the inverse of the bath Green matrix as
\begin{displaymath}
 \Gbath^{-1}(\omega)=\mathbf{\Sigma}_c(\omega)+\left(\int d\ksmall\; 
    \Big(\omega+\mu-\mathbf{H}(\ksmall)-\boldsymbol{\Sigma}_c(\omega)\Big)^{-1}\right)^{-1}.
\end{displaymath}
Considering the limit $\omega\to\infty$, expanding to order $1/\omega^2$, using (\ref{Hc}), and comparing to (\ref{diagbath}) we find
\begin{equation}\label{sumrule}
 \sum_l \mathbf{V}_l\,\mathbf{V}_l^\dagger = \int d\ksmall\,\mathbf{H}^2(\ksmall)
                                - \left(\int d\ksmall\,\mathbf{H}(\ksmall)\right)^2\,.
 %                                      - \mathbf{H}_c^2\;. 
\end{equation}
To illustrate this hybridization sum-rule we consider a representative set of examples.

\subsection{Single site}
We consider a $d$-dimensional lattice with hoppings $t_n$ to the $z_n$ $n^{th}$-nearest neighbors. For $N_c=1$ we have $\mathbf{H}(\mathbf{k})=\varepsilon_\mathbf{k}$. Thus we find for the hybridizations 
\begin{equation}\label{singlesite}
 \sum_l V^2_l
 = \frac{1}{(2\pi)^d}\int_{-\pi}^\pi d^d\mathbf{k}\,\varepsilon_\mathbf{k}^2
 = \sum_n z_n\,t_n^2\,,
\end{equation}
where the integral is just the second moment of the density of states, so that the last equation follows as in the recursion method.\cite{recursion} 
For a Bethe lattice of connectivity $z$ with hopping matrix element $t/\sqrt{z}$ the sum-rule reduces to $\sum_l V^2_l = t^2$. 
% how about multi-orbital?

\subsection{CDMFT}
We start by considering a linear chain with nearest neighbor hopping $t$ and a three-site cluster $N_c=3$. In the CDMFT gauge we have 
\begin{equation}
 \mathbf{H}(\tilde{k})=-t\,\left(\begin{array}{lcr}
                           0               & 1\; & e^{-3i\tilde{k}}\\
                           1               & 0   & 1\\
                           e^{3i\tilde{k}} & 1   & 0 
                         \end{array}\right)
\end{equation}
so that $\mathbf{H}_c$ is the original single-electron Hamiltonian restricted to the cluster:
\begin{equation}
 \mathbf{H}_c = \frac{3}{2\pi}\int_{-\pi/3}^{\pi/3} d\tilde{k}\,\mathbf{H}(\tilde{k})
 = -t\,\left(\begin{array}{ccc} 0&1&0\\ 1&0&1\\ 0&1&0\end{array}\right)\;.
\end{equation}
The sum-rule (\ref{sumrule}) then is
\begin{equation}
 \left(\sum_l V_{l,i} \bar{V}_{l,j}\right) 
 = \left(\begin{array}{lll} t^2&0&0\\ 0&0&0\\ 0&0&t^2\end{array}\right)\;,
\end{equation}
i.e., only the sites on the surface of the cluster couple to the bath. 
%If we allow also second nearest neighbor hopping with matrix element $t''$, we find
%\begin{equation}
% \left(\sum_l \bar{V}_{l,\mu} V_{l,\nu}\right) =
% \left(\begin{array}{ccc}
%  t^2+t''^2 & t t'' & 0\\
%  t t'' & 2 t''^2 & t t''\\
%  0 & t t'' & t^2+t''^2
% \end{array}\right)\;.
%\end{equation}
\begin{figure}
\center
\includegraphics[width=3in]{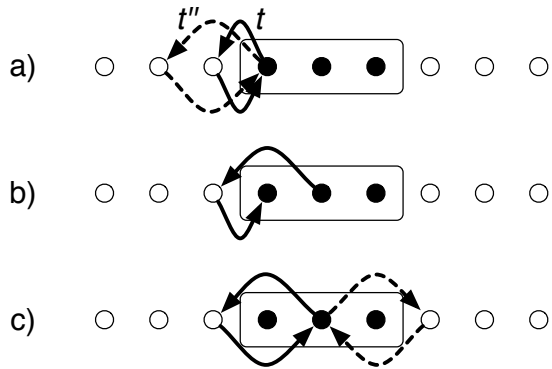}
\caption{CDMFT sum-rules for a one-dimensional 3-site cluster and nearest and next-nearest neighbor hoppings $t$ and $t''$, respectively: 
 a) $\sum_l |V_{l,1}|^2=t^2+t''^2$, b) $\sum_l \bar{V}_{l,1}V_{l,2}=t\,t''$, and 
 c) $\sum_l |V_{l,2}|^2=2t''^2$. The hybridizations are given by the two-step hopping processes that are lost when cutting the cluster out of the original lattice.}
\label{sumrule3}
\end{figure}

% second moment H^2
The general CDMFT hybridization sum-rule (\ref{sumrule}) can be easily visualized: The integral over the Brillouin-zone of the cluster projects the single-electron part of the full Hamiltonian to the cluster (see eqn.~(\ref{Hc})). The matrix elements of $\mathbf{H}_c^2$ are thus the two-step hoppings that are possible on the cluster. Likewise the integral over the Hamiltonian squared gives the second moments, only that here the  intermediate site is not restricted to the cluster. Thus the sum-rule matrix is given by the second-order paths between cluster-sites that proceed via a site outside the cluster. This is illustrated in figure \ref{sumrule3}. As a special case, for a single site we obtain the second equality in (\ref{singlesite}).

The vanishing of a matrix element in the sum-rule only implies that the corresponding matrix element of the bath Green matrix decays faster than $1/\omega$ for large $\omega$. For a diagonal element, however, all terms in $\sum_l V_{l,i}\bar{V}_{l,i}$ are positive. Thus a vanishing sum means that all terms must be zero. Hence the sum-rule implies that cluster-sites that are so far in the interior that they cannot be reached by hopping from outside the cluster do not couple to bath and that all matrix elements of the bath Green function involving such a site $i$ are given by $\Gb_{ij}(\omega)=\omega+\mu-(\mathbf{H}_c)_{ij}$ for all $\omega$. In that sense the bath hybridizes only to the surface of the cluster and we see that the hybridization-strength to these sites does not decrease for increasing cluster size $N_c$. 
% and is independent of the interaction $U$.

\subsection{DCA}
We start again by considering the 3-site cluster.
In the DCA gauge we write 
\begin{equation}
 \mathbf{H}(\tilde{k})=-t\,\left(\begin{array}{lcr}
                    0 & e^{i\tilde{k}} & e^{-i\tilde{k}}\\
                    e^{-i\tilde{k}} & 0 & e^{i\tilde{k}}\\
                    e^{i\tilde{k}} & e^{-i\tilde{k}} & 0
                   \end{array}\right)\;.
\end{equation}
Now $\mathbf{H_c}$ has translation symmetry, but the hopping matrix element is rescaled by $\sin(\pi/N_c)/(\pi/N_c)$:
\begin{equation}
 \mathbf{H}_c = \frac{3}{2\pi}\int_{-\pi/3}^{\pi/3} d\tilde{k}\,\mathbf{H}(\tilde{k})
 = -\frac{3\sqrt{3}}{2\pi}\,t\,
     \left(\begin{array}{ccc} 0&1&1\\ 1&0&1\\ 1&1&0\end{array}\right)\;.
\end{equation}
Since all matrices in (\ref{sumrule}) are periodic, it is convenient to transform to $k$-space. With $V_{l,K}= \sum_i V_{l,i}e^{iKr_i}/\sqrt{N_c}$ and the coarse-graining factor $\tau=3\sqrt{3}/2\pi$ we find 
\begin{eqnarray*}
 \sum_l |V_{l,K=0}|^2 &=& (2+\tau-4\tau^2)t^2\\
 \sum_l |V_{l,K=\pm2\pi/3}|^2 &=& (2-\tau/2-\tau^2)t^2\;.
\end{eqnarray*}
%$|V_{l,K=0}|^2=(4\pi^2+3\sqrt{3}\pi-54)t^2/2\pi^2$ and $|V_{l,K=\pm2\pi/3}|^2=(8\pi^2-3\sqrt{3}\pi-27)t^2/4\pi^2$.

The hybridization sum-rule (\ref{sumrule}) is then, likewise, diagonal in the cluster-momenta $\mathbf{K}$
\begin{equation}\label{DCAsum}
 \sum_l |V_{l,\mathbf{K}}|^2 = \int d\ksmall\,\varepsilon_{\mathbf{K}+\ksmall}^2
            -\left(\int d\ksmall\,\varepsilon_{\mathbf{K}+\ksmall} \right)^2\,,
\end{equation}
while all terms $V_{l,\mathbf{K}}\bar{V}_{l,\mathbf{K}'}$ mixing different cluster momenta vanish.
As a special case, for a single site the above sum-rule is just the first equality in (\ref{singlesite}). 
Expanding $\varepsilon_{\mathbf{K}+\mathbf{k}}$ around $\mathbf{K}$, we find that for a $d$-dimensional system $\sum_l |V_{l,\mathbf{K}}|^2$ decreases with cluster size as $1/N_c^{2/d}$, while all cluster-sites couple with the same strength to the bath.\cite{Maier00}

\subsection{Discussion}
From the sum-rules we recover\cite{review05} that the individual hybridizations in CDMFT are independent of cluster size, while for DCA they decrease with cluster size as $N_c^{-2/d}$.  Interestingly this means that for a $d$-dimensional system in CDMFT the overall coupling to the bath scales as $N_c^{(d-1)/d}$, while in DCA it scales as $N_c^{(d-2)/d}$. For non-local properties a DCA calculation is therefore expected to converge faster with cluster size.\cite{comment} For a calculation where we represent the bath by discrete degrees of freedom this decrease in hybridization strength does, however, not help very much as we still need bath-sites to fit the hybridizations, even if they are small. With increasing DCA cluster size we thus have to parametrize $N_c$ baths, one for each $\mathbf{K}$. In CDMFT the situation is more fortunate, as the sum-rules imply that many hybridizations vanish and we only need to parametrize the coupling of surface-sites to the bath. 
% In the atomic limit ($t\to0$) the hybridizations for both methods vanish.

The lack of translational invariance in CDMFT has two important practical implications. First, the full Green matrix has to be calculated, instead of just its diagonal. 
% As an example, for linear clusters, assuming inversion symmetry, we need to calculate $\lceil N_c/2\rceil$ elements [should be Nc!?] of the Green matrix in DCA, while in CDMFT we have to determine an additional $\sum_{k=1}^{\lfloor N_c/2\rfloor} (N_c-2k+1)$ different off-diagonal matrix elements. 
Second, when calculating local quantities, like the density per site, in CDMFT we have a choice of inequivalent sites, or we could consider the average over all sites. In a gapped system the best choice is the innermost site, \cite{reply} however, in such a situation it might be better to do a straight Lanczos calculation with $N_c+N_b$ cluster sites, instead of using $N_b$ bath sites.\cite{DM}

\section{Symmetries}\label{SecSymmetry}
In the absence of spontaneous symmetry breaking the symmetries of the cluster (point-symmetries in CDMFT and additionally translation symmetry in DCA) are reflected in the Green matrix. In a symmetry broken state with long-range order, like an antiferromagnet or a charge-density wave, the symmetry of the Green matrix is accordingly lowered. To exploit this symmetry we introduce vectors on the cluster that transform according to its irreducible representations. We write these vectors as $\mathbf{w}_{I,\nu}$ where $I$ is the irreducible representation and $\nu=1\ldots N_I$ counts linear independent vectors transforming according to $I$. On an $N_c$-site cluster we can choose $N_c$ such vectors that are orthonormal. Defining the matrix $\mathbf{W}=(\mathbf{w}_{I,\nu})$ of these vectors, we can block-diagonalize the bath Green matrix: $\mathbf{W}^\dagger \Gbath^{-1}\mathbf{W}$ has blocks of dimension $N_I$ corresponding to the different irreducible representations $I$. Since $\mathbf{W}^\dagger \Gbath^{-1}\mathbf{W}$ is block diagonal for all $\omega$, it follows from equation (\ref{diagbath}), that $\mathbf{W}$ must also block-diagonalize the individual hybridization matrices $\mathbf{V}_l\,\mathbf{V}_l^\dagger$. Therefore the hybridization vectors must transform according to an irreducible representation:
% if it was a linear combination of several irreps, we would get non-zero elements outside blocks (in I,J sector); show for block-diag form of hybridization
They can be written as $\mathbf{V}_{l}=\sum_\nu V_{l;I,\nu}\,\mathbf{w}_{I,\nu}$ for some irreducible representation $I$. If the $\mathbf{V}_l$ also had components $\mathbf{w}_{J,\nu}$ of a different irreducible representation $J\ne I$ this would produce a hybridization matrix that could not be block-diagonalized. 
% This can only happen for bath-sites with identical energy $\varepsilon_l$: Assume $\mathbf{V}_l$ and $\mathbf{V}_{l'}$ are the hybridizations for two bath-sites with $\varepsilon_l=\varepsilon_{l'}$. Then we can form arbitrary linear combinations of the hybridization matrices and hence of the hybridization vectors. For all these linear combinations the sum of the hybridization matrices must be block diagonal, and hence we can choose the hybridization vectors such that they transform according to irreducible representations.

We thus find that the bath-sites can be arranged into sets corresponding to the different irreducible representations. For fitting a block of the symmetrized bath Green matrix we need then only consider bath-sites of the respective irreducible representation. If the block is one-dimensional we can choose the corresponding hybridization real. 
An early example is the bonding-antibonding transformation introduced in reference \onlinecite{mazurenko}.

%Sectors corresponding to different irreducible representation are only coupled through the Hubbard interaction $U$ when solving the Anderson impurity model. We note that the coupling to bath-sites corresponding to an irreducible representation other than the unit representation lowers the symmetry of the impurity Hamiltonian with respect to that of the Green matrix.

\subsection{CDMFT}
\begin{figure}
 \center
 \includegraphics[width=2.8in]{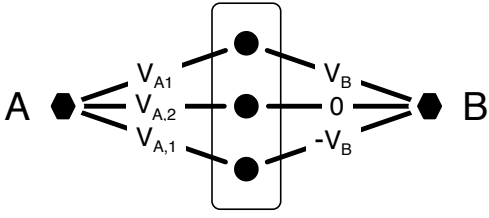}
 \caption{Hybridization of bath-sites of symmetry $A$ and $B$ to a 3-site cluster. As defined in table \ref{irreps}, $A$ is the unit representation, so a bath-site of type $A$ has the same hybridization $V$ to all cluster-sites that are equivalent by symmetry. $B$ is the antisymmetric representation, so the hybridization of a bath-site of type $B$ to cluster-sites that are related by mirror symmetry have the opposite sign. Consequently the hybridization to the central site of a linear cluster with an odd number of sites vanishes in the $B$ representation.}
 \label{bath3}
\end{figure}

As an example we consider a linear cluster of 3 sites as shown in figure \ref{bath3}. The symmetry is $C_2$ (see table \ref{irreps}). Transforming to the basis vectors $\mathbf{w}_{A,1}=(|1\rangle+|3\rangle)/\sqrt{2}$ and $\mathbf{w}_{A,2}=|2\rangle$ of symmetry $A$ (see table \ref{irreps}) and $\mathbf{w}_B=(|1\rangle-|3\rangle)/\sqrt{2}$, we find the transformed bath Green matrix
\begin{displaymath}%\label{Gblock}
% > W:=[[1,0,1],[0,sqrt(2),0],[1,0,-1]]/sqrt(2);
% > G:=[[G11,G12,G13],[G21,G22,G21],[G13,G12,G11]];
% G31 -> G13; G33 -> G11
% > evalm(transpose(W)&*G&*W): map(simplify, %);
 \mathbf{W}^\dagger\Gbath^{-1}\mathbf{W}=\left(\begin{array}{ccc}
         \Gb_{11}+\Gb_{13}    & \sqrt{2}\,\Gb_{12} & 0\\
         \sqrt{2}\,\Gb_{21} & \Gb_{22}           & 0\\
          0               & 0                &\Gb_{11}-\Gb_{13}
        \end{array}\right)\,.
\end{displaymath}
A bath-site of irreducible representation $A$ contributes to the first block and has the same hybridization $V_{A,1}$ to the outer cluster-sites plus an independent hybridization parameter $V_{A,2}$ to the central site. A bath-site of irreducible representation $B$ contributes to the second block. For such a bath-site the hybridization to cluster-sites that are related by mirror symmetry have opposite signs. Consequently, the hybridization to the central site vanishes.

\begin{table}
\begin{minipage}{18ex}
 $\begin{array}{c|rr}
  C_2 & E & \sigma_v \\\hline
  A   & 1 &  1\\
  B   & 1 & -1
 \end{array}$\\[2ex]
% $\begin{array}{c|rrrr}
%  C_{2v} & E & C_2 & \sigma_v & \sigma_v'\\\hline
%  A_1    & 1 &  1 &  1 &  1\\
%  A_2    & 1 &  1 & -1 & -1\\
%  B_1    & 1 & -1 &  1 & -1\\
%  B_2    & 1 & -1 & -1 &  1
% \end{array}$\\[2ex]
 $\begin{array}{c|rrrrr}
  C_{3v} & E & 2C_3 & 3\sigma_v\\\hline
  A_1    & 1 &  1   &  1\\
  A_2    & 1 &  1   & -1\\
  E      & 2 & -1   &  0
 \end{array}$
\end{minipage}\hspace{6ex}
 $\begin{array}{c|rrrrr}
  C_{4v} & E & 2C_4 & C_4^2 & 2\sigma_v & 2\sigma_d\\\hline
  A_1    & 1 &  1 &  1 &  1 &  1\\
  A_2    & 1 &  1 &  1 & -1 & -1\\
  B_1    & 1 & -1 &  1 &  1 & -1\\
  B_2    & 1 & -1 &  1 & -1 &  1\\
  E      & 2 &  0 & -2 &  0 &  0
 \end{array}$
 \caption{Character tables of the point groups $C_{1v}$, $C_{2v}$, $C_{3v}$, and $C_{4v}$.}
 \label{irreps}
\end{table}

\begin{figure}
 \center
 \includegraphics[width=2.3in]{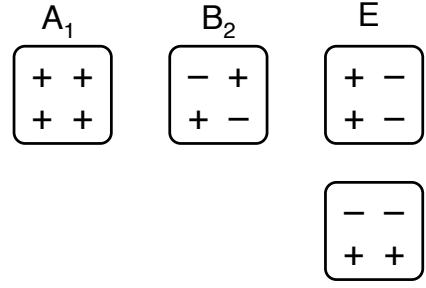}
 \caption{Hybridization of bath-sites of symmetry $A_1$, $B_2$, and $E$ to a $2\times2$ cluster. For a given irreducible representation the absolute value of the hybridization to all cluster-sites is the same, while the signs are indicated in the figure. Non-trivial hybridizations corresponding to irreducible representations $A_2$ or $B_1$ only appear for larger clusters.}
 \label{twobytwo}
\end{figure}

The situation is slightly more complicated when the symmetry group has irreducible representations of dimension higher than one. The simplest example is the $2\times2$ cluster with $C_{4v}$ symmetry. With $\mathbf{w}_{A_1}=(|1\rangle+|2\rangle+|3\rangle+|4\rangle)/2$, $\mathbf{w}_{B_2}=(|1\rangle-|2\rangle+|3\rangle-|4\rangle)/2$, and the pair $\mathbf{w}_{E,1}=(|1\rangle-|2\rangle-|3\rangle+|4\rangle)/2$, $\mathbf{w}_{E,2}=(|1\rangle+|2\rangle-|3\rangle-|4\rangle)/2$ we find that $\mathbf{W}^\dagger\Gbath^{-1}\mathbf{W}$ is diagonal with diagonal elements
\begin{eqnarray*}
% > W:=[[1,1,1,1],[1,-1,-1,1],[1,1,-1,-1],[1,-1,1,-1]]/2;
% > G:=[[G11,G12,G13,G12],[G12,G11,G12,G13],[G13,G12,G11,G12],[G12,G13,G12,G11]];
% > evalm(transpose(W)&*G&*W): map(simplify, %);
 (\mathbf{W}^\dagger\Gbath^{-1}\mathbf{W})_{11} &=& \Gb_{11}+2\Gb_{12}+\Gb_{13}\\
 (\mathbf{W}^\dagger\Gbath^{-1}\mathbf{W})_{22} &=& \Gb_{11}-2\Gb_{12}+\Gb_{13}\\
 (\mathbf{W}^\dagger\Gbath^{-1}\mathbf{W})_{33} &=& \Gb_{11}-\Gb_{13}\\
 (\mathbf{W}^\dagger\Gbath^{-1}\mathbf{W})_{44} &=& \Gb_{11}-\Gb_{13}\\
\end{eqnarray*}
A bath-site of symmetry $A_1$ has the same hybridization to all cluster-sites while for a bath-site of symmetry $B_2$ the hybridizations have alternating signs: $\mathbf{V}_l^\dagger=\bar{V}_l\,(1,-1,1,-1)$. To realize the two-dimensional representation $E$ we need two bath-sites $l_1$ and $l_2$ with degenerate energies $\varepsilon_{l_1}=\varepsilon_{l_2}=\varepsilon_l$ and hybridizations: $\mathbf{V}_{l_1}^\dagger=\bar{V}_l\,(1,-1,-1,1)$ and $\mathbf{V}_{l_2}^\dagger=\bar{V}_l\,(1,1,-1,-1)$. This is illustrated in figure \ref{twobytwo}.

\subsection{DCA}
As an example for DCA we consider a 3-site cluster with periodic boundary conditions. The symmetry group is $C_{3v}$ (translations and inversion). Hence we introduce the basis vector $\mathbf{w}_{A_1}=(|1\rangle+|2\rangle+|3\rangle)/\sqrt{3}$, corresponding to $k=0$, while the vectors formed by $\sin(2\pi/3)$ and $\cos(2\pi/3)$ give the $E$ representation: $\mathbf{w}_{E,1}=(|1\rangle-|2\rangle)/\sqrt{2}$ and $\mathbf{w}_{E,2}=(|1\rangle+|2\rangle-2|3\rangle)/\sqrt{6}$.
\begin{displaymath}
% > W:=[[1/sqrt(3), 1/sqrt(2), 1/sqrt(6)], \                                       
% >     [1/sqrt(3),-1/sqrt(2), 1/sqrt(6)], \
% >     [1/sqrt(3), 0        ,-2/sqrt(6)]];
% > G:=[[G11,G12,G12], [G12,G11,G12], [G12,G12,G11]];
% > evalm(transpose(W)&*G&*W): map(simplify, %);
 \mathbf{W}^\dagger\Gbath^{-1}\mathbf{W}=
 \left(\begin{array}{ccc}
  \Gb_{11}+2\Gb_{12} & 0 & 0\\
  0 & \Gb_{11}- \Gb_{12} & 0\\
  0 & 0 & \Gb_{11}- \Gb_{12}\\
 \end{array}\right)\,.
\end{displaymath}
In general bath-sites corresponding to the gamma point have the same hybridization to all cluster-sites, while those corresponding to $k=\pi$ have alternating hybridizations. For all other $k$-points we need two degenerate bath-sites, with hybridizations $V_{l_1,\mu}=V_l\,\sin(k\mu)$ and $V_{l_2,\mu}=V_l\,\cos(k\mu)$ to cluster-site $\mu$.

\subsection{Cluster replica}
\begin{figure}
\center
\includegraphics[width=1.5in]{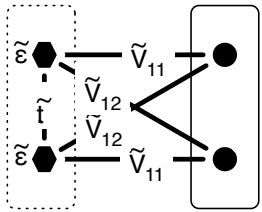}
 \caption{Replica of a 2-site cluster in the bath. Diagonalizing the Hamiltonian for the two bath-sites leads to a bath-site of type $A$ (even representation) with energy $\varepsilon_A=\tilde{\varepsilon}-\tilde{t}$ and hybridization $V_A=(\tilde{V}_{11}+\tilde{V}_{12})/\sqrt{2}$ and a site of type $B$ with $\varepsilon_B=\tilde{\varepsilon}+t$ and $V_B=(\tilde{V}_{11}-\tilde{V}_{12})/\sqrt{2}$.}
 \label{replica2}
\end{figure}

Instead of implementing the symmetry of the Green matrix as described above, one might construct the bath out of replica of the $N_c$-site clusters.\cite{civelliPhD} For a two-site cluster this means that bath-sites come in pairs, with on-site energy $\tilde{\varepsilon}$, hopping $-\tilde{t}$ between the bath-sites, and hybridization $\tilde{V}_{11}$ and $\tilde{V}_{12}$ to the cluster as illustrated in figure \ref{replica2}. Diagonalizing such a bath-pair, we obtain one bath-site of symmetry $A$ with on-site energy $\varepsilon_A=\tilde{\varepsilon}-\tilde{t}$ and hybridization $V_A=(\tilde{V}_{11}+\tilde{V}_{12})/\sqrt{2}$ and one bath-site of symmetry $B$ with $\varepsilon_B=\tilde{\varepsilon}+t$ and $V_B=(\tilde{V}_{11}-\tilde{V}_{12})/\sqrt{2}$.

To generalize this approach, let $\tilde{H}_b$ be the Hamiltonian for an $N_c$-site cluster with general on-site energies and hoppings that respect the symmetry of the Green matrix. Furthermore let bath-site $n$ hybridize to cluster-site $i$ with $\tilde{V}_{ni}$. These hybridizations are chosen symmetric under simultaneous symmetry transformations of the original cluster and the bath replica. Diagonalizing $\tilde{H}_b$ we obtain $N_c$ eigenstates $\mathbf{\varphi}_l$ with energy $\varepsilon_l$. These can be considered as bath-sites that hybridize to cluster-site $i$ with $V_{l,i}=\sum_n \varphi_{l,n}\,\tilde{V}_{n,i}$.

Being the eigenstates of the Hamiltonian $\tilde{H}_b$, the $\mathbf{\varphi}_l$ transform according to the irreducible representations of the symmetry of the Green matrix.\cite{accdeg} Therefore they can be written as linear combination $\mathbf{\varphi}_l=\sum_\nu \alpha_{l;I,\nu} \mathbf{w}_{I,\nu}$ for some irreducible representation $I$. From this we can conclude that a cluster replica gives rise to $N_I$ bath-sites of symmetry $I$. By working with cluster replica we thus sacrifice the freedom of choosing the irreducible representations for the bath individually. Moreover, it is not straightforward to find a proper parametrization. $\tilde{H}_b$ must be chosen such that all accidental degeneracies can be lifted. For symmetries with higher-dimensional irreducible representations there will be, however, corresponding essential degeneracies. 
% A practical prescription is to include in $\tilde{H}_b$ all different hoppings allowed by symmetry and an energy-shift $\tilde{\varepsilon}$. After diagonalization of $\tilde{H}_b$, the hybridization of bath-site $l$ is given by $V_{l,i}=\sum_\nu \alpha_{l;I,\nu}\,(\sum_n w_{I,\nu;n}\tilde{V}_{n,i})=\sum_\nu \alpha_{l;I,\nu}\,\tilde{w}_{I,\nu;i}$. 
Moreover, working with cluster replica, we cannot fit the individual blocks of the bath Green matrix with the minimal set of symmetry-adapted parameters, but have to solve the optimization problem for the full bath Green matrix and all parameters. 
Thus, using cluster replica is less flexible than using individual irreducible representations and it leads to a more complicated fitting procedure, in particular when considering large baths.

\section{The Hubbard chain}\label{SecEsperti}
We now discuss the one-dimensional Hubbard model, for which exact results are available from the Bethe ansatz\cite{Bethe} and which has been studied with CDMFT using Lanczos\cite{bolech,esperti,civelliPhD} and QMC,\cite{kyungQMC} as well as with the variational cluster approximation.\cite{VCA}  Here going from single-site DMFT to a cluster description makes a qualitative difference: For a paramagnetic single-site calculation antiferromagnetism is completely suppressed, while on a cluster we will have short-ranged antiferromagnetic correlations, even if we impose a paramagnetic bath.

\begin{figure}
 \center
 \includegraphics[width=3in]{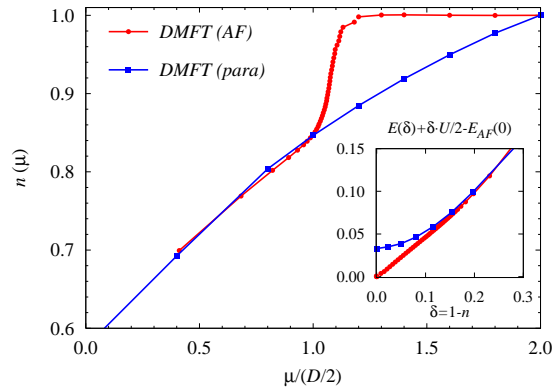}
 \caption{\label{AFsinglesite}
  (Color online) Density as a function of chemical potential for a Hubbard model on the Bethe lattice with half band width $D$ and $U=2\,D$ in single-site DMFT with paramagnetic (para) and antiferromagnetic (AF) bath (blue squares and red circles respectively). The AF curve comprises a left and a right branch: the left one has been obtained by fixing the magnetization from 0.01 to 0.82 (from left to right) and determining the chemical potential. The right branch has been obtained by just decreasing the chemical potential from the half-filling value ($\mu=U/2$) down to the point where convergence is no longer found ($\mu=0.6$). The magnetization in the right branch is about 0.85. In the inset we compare the total energy of the AF and para phases as a function of doping. The AF phase is the stable one up to about $\delta=0.2$. We have added $\delta\,U/2-E_{AF}(0)$ to $E(\delta)$ to allow better comparison with Fig.~1 of Ref.~\onlinecite{vanDongen} and Fig.~8 of Ref.~\onlinecite{Zitzler}. 
  %One last remark about the AF curve (left branch): this has been obtained with fixed-magnetization runs forcing the calculation in a particular sector. This is the only way in which we manage to get converged solutions. Fixed chemical potential runs suffer from slow periodic density oscillations and fixed-magnetization runs allowing for several sectors don't converge. At fixed magnetization we tried different (fixed) sectors and then we chose the one which gives the lowest total energy (lattice kinetic energy $+ U\langle n_\uparrow n_\downarrow\rangle$).
  }
\end{figure}

This inclusion of antiferromagnetic correlations might well be the cause for the spectacular difference between the single-site and 2-site CDMFT calculations of the density as a function chemical potential reported in reference \onlinecite{esperti}. For illustration, in Fig.~\ref{AFsinglesite}, we compare the density as a function of the chemical potential for a single-site calculation with paramagnetic and antiferromagnetic bath.\cite{AF} 
They represent two limiting cases, the former being always metallic, the latter yielding a gap, which is overestimated as we are using an antiferromagnetic bath to mimic the short-ranged correlations present in the Hubbard chain.

\subsection{Bath Green matrix}
\begin{figure}
 \center
 \includegraphics[width=3in]{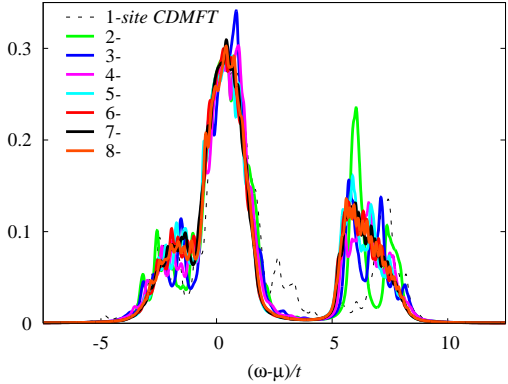}
 \caption{\label{CDMFTbath}
  (Color online) Plot of the CDMFT bath Green function $\mathcal{G}_{11}^{-1}(\omega)-(\omega+\mu)$ on the real axis for linear clusters with nearest-neighbor hopping $t$. Increasing the size of the cluster the hybridization hardly changes. The plots show calculations for chains with $U=6t$ and $\mu=0.5t$.}
\end{figure}

% electron-hole symmetry?
For a one-dimensional lattice the CDMFT bath Green matrix simplifies drastically: When removing the $N_c$-site cluster from the lattice, we are left with two disconnected pieces. Thus a vanishing sum-rule (\ref{sumrule}) in one dimension means that the hybridization matrix vanishes for all frequencies. For the irreducible representations of the bath this means that bath-sites of symmetry $A$ and $B$ come in degenerate pairs: $\varepsilon_{A_l}=\varepsilon_{B_l}$ and $V_{A_l}=V_{B_l}$. Thus for the Hubbard chain with nearest neighbor hopping only the outer-most cluster sites hybridize with the bath and the bath parametrizations are identical by symmetry.\cite{civelliPhD} The evolution of the bath Green matrix element ${\cal G}^{-1}_{11}$ with cluster size is shown in Fig.~\ref{CDMFTbath}. We find that the bath Green matrix elements hardly depend on cluster size and even for a single-site calculation the bath is already similar to that for a large cluster.
%OLD: If we consider CDMFT for a cluster with only nearest-neighbor hopping we know from the hybridization sum-rules (\ref{sumrule}) that only sites on the surface couple to the bath. For a one-dimensional cluster of $N_c$ sites, the only non-vanishing elements of the bath Green matrix are $(\mathbf{W}^\dagger\Gbath^{-1}\mathbf{W})_{11}$ and $(\mathbf{W}^\dagger\Gbath^{-1}\mathbf{W})_{N_cN_c}$ corresponding to $\mathbf{w}_A=(|1\rangle+|N_c\rangle)/\sqrt{2}$ and $\mathbf{w}_B=(|1\rangle-|N_c\rangle)/\sqrt{2}$. Since the bath of symmetry $A$ and that of symmetry $B$ have to fit the same function, $\Gb_{11}$, the bath-sites must be degenerate, $\varepsilon_{A_l}=\varepsilon_{B_l}$ and $V_{A_l}=V_{B_l}$. This degeneracy has been observed empirically in reference \onlinecite{esperti}. Transforming these pairs of degenerate bath-sites we obtain one set of bath-sites that only couple to cluster-site 1 and a separate set with identical parameters coupling only to cluster-site $N_c$. The corresponding bath Green function $\Gb_{11}$ is shown in figure \ref{CDMFTbath}.

\begin{figure}
 \center
 \includegraphics[width=2.7in]{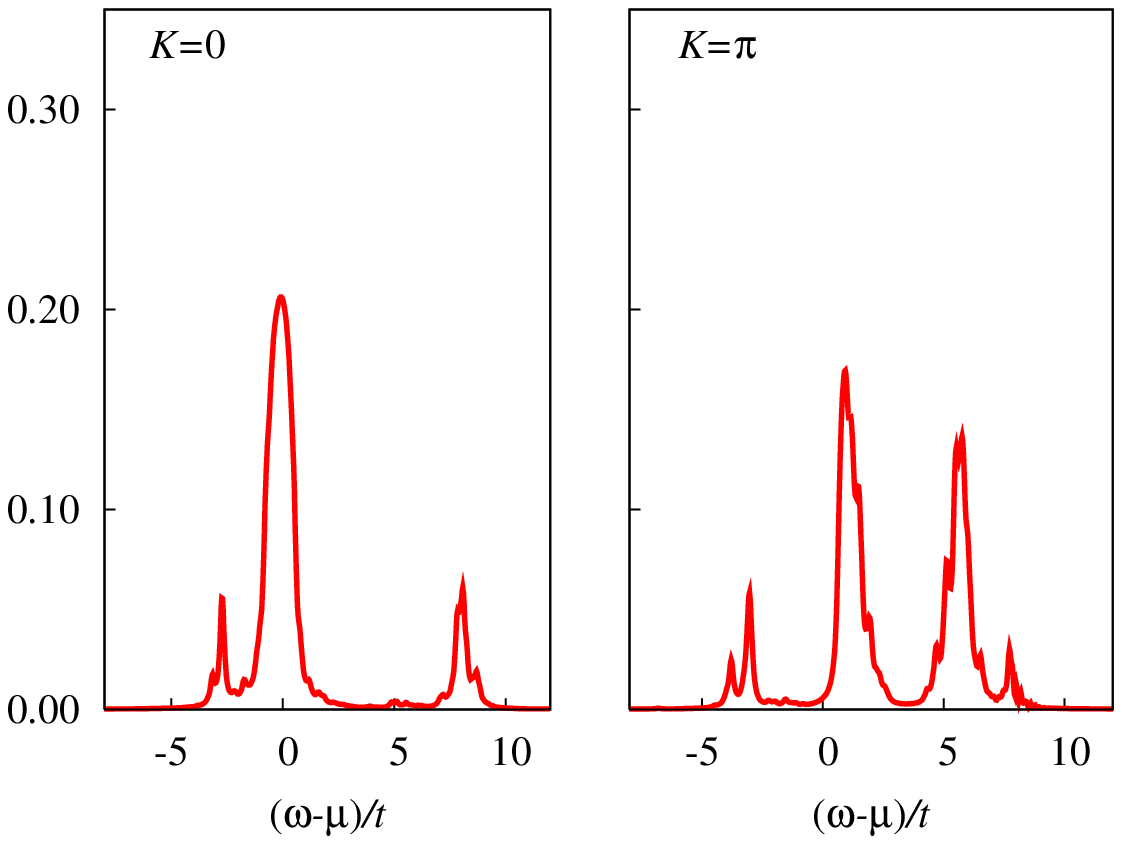}
 \includegraphics[width=2.7in]{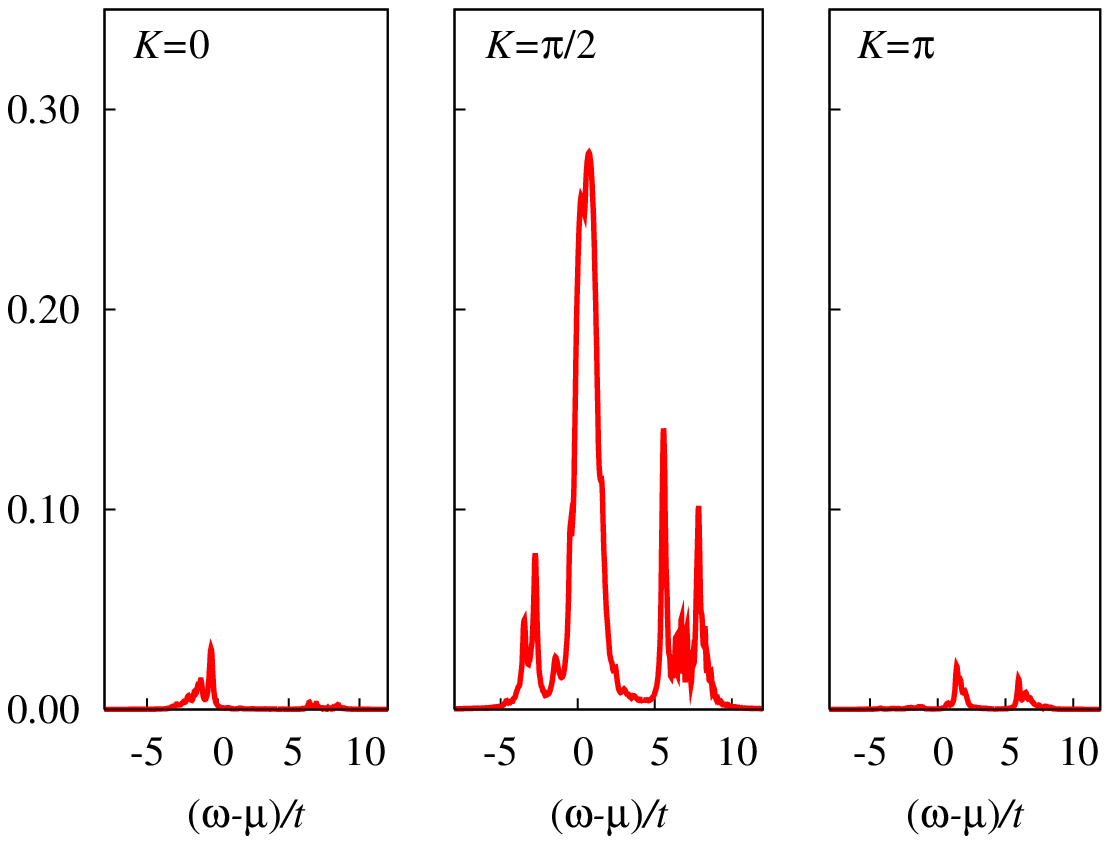}
 \includegraphics[width=2.7in]{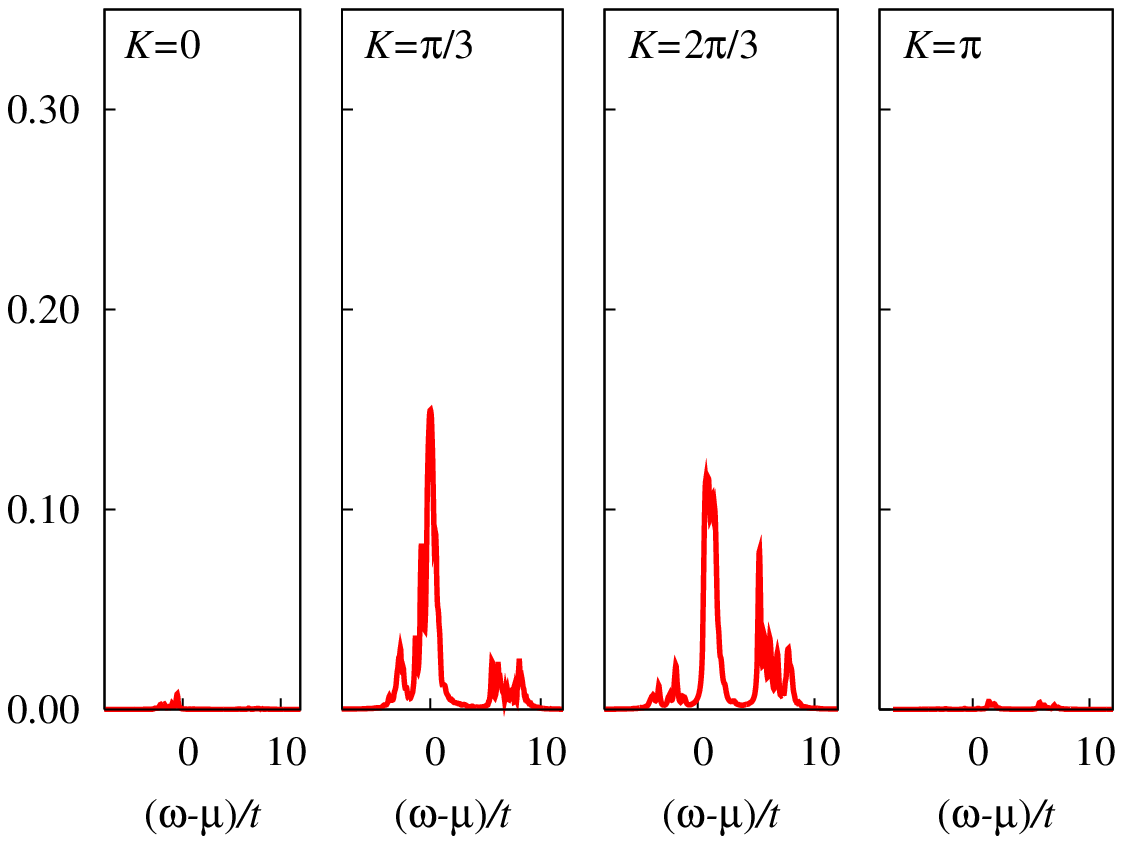}
 \includegraphics[width=2.7in]{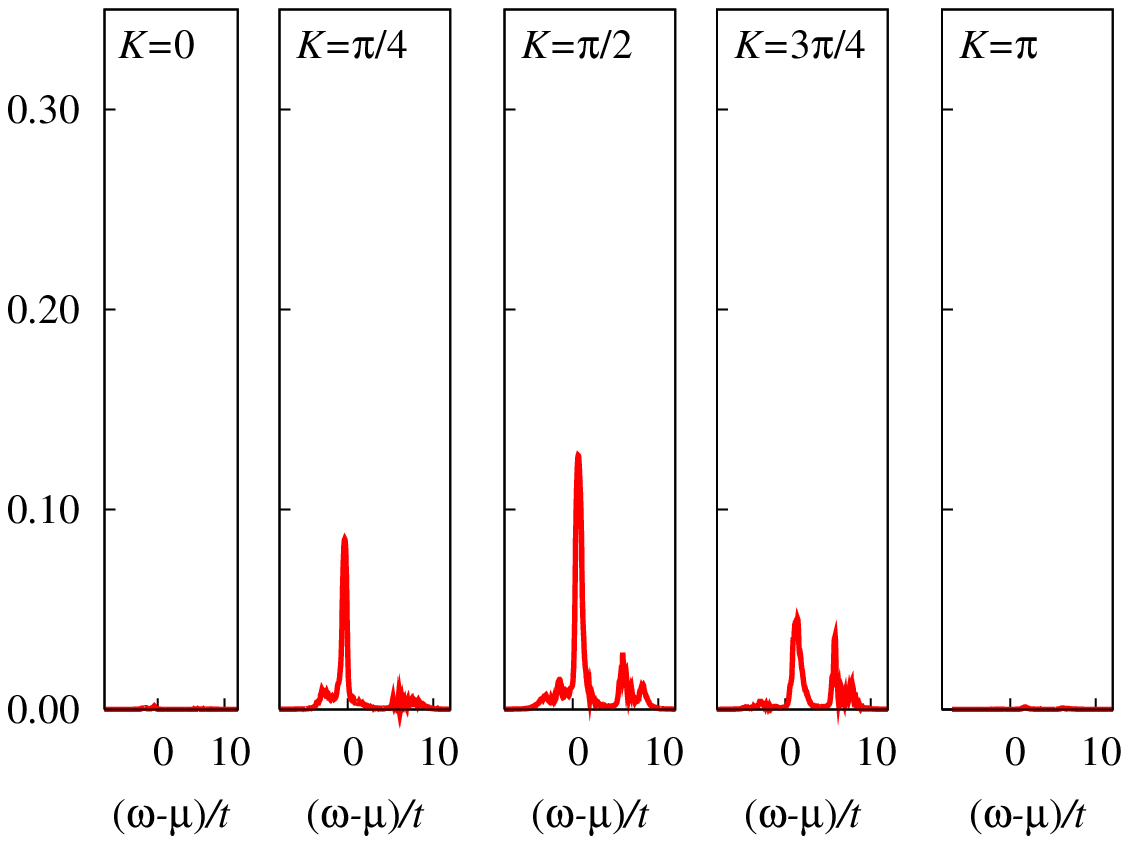}
 \caption{\label{DCAbath}
  (Color online) Plot of the DCA bath Green functions $\mathcal{G}^{-1}_{K}(\omega)-(\omega + \mu-\int d\tilde{k}\,\varepsilon_{K+\tilde{k}})$ on the real axis for linear clusters with nearest-neighbor hopping $t$. Increasing  cluster-size ($N_c=$2, 4, 6, and 8 from top to bottom) the number of independent functions to fit increases. Note that $K=0$ and $K=\pi$ are associated with single bath-sites, while all the other $K$-points need to be described by pairs of bath-sites. From the figures it is clear how the total spectral weight decreases with increasing $N_c$. The plots show calculations for chains with $U=6t$ and $\mu=0.5t$.}
\end{figure}

In contrast, in the DCA we get a non-vanishing hybridization for each $K$-point of the cluster. This is shown in figure \ref{DCAbath}. While the hybridization strength per $K$-point decreases with cluster size, we still have to parametrize all of them, possibly except for $K=0$ and $\pi$, which almost vanish already for moderate cluster sizes.
%since DCA requires too many bath-sites, in the following we mainly focus on CDMFT.
%coarse-grained band energy
%\begin{equation}
% \bar{\varepsilon}_K = \int d\tilde{k}\,\varepsilon_{K+\tilde{k}}
%\end{equation}

\subsection{Fitting the bath Green matrix}
\label{fitting}
The most critical step in calculations with a finite bath is the determination of the parameters for the impurity Hamiltonian (\ref{Hand}). This is usually done by fitting the bath Green function on the imaginary axis: A fictitious temperature $1/\beta$ is introduced and the sum of the squared difference between $\Gbath$ and its parametrized version (\ref{diagbath}) over the Matsubara frequencies up to some cutoff is minimized.\cite{DMFTreview} This procedure is fairly robust if the number of bath-sites is sufficiently large. For cluster calculations the number of effective bath-sites per fitted bath Green function can, however, be quite small. In such a situation details of the fitting procedure are important and are accordingly discussed in the literature.\cite{bolech,esperti,kyung06,civelliPhD,civelli07,imada07}

\begin{figure}
 \center
 \includegraphics[width=3in]{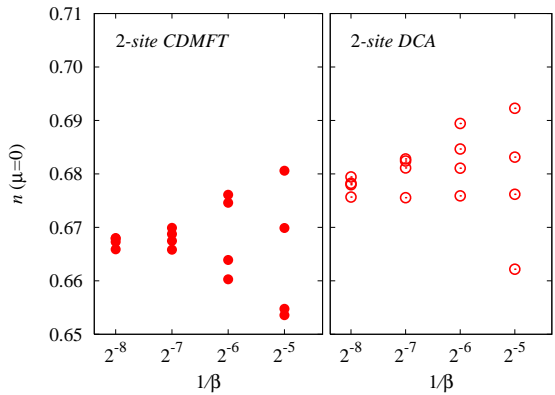}
 \caption{\label{spread_all}
  (Color online) Density for a 2-site cluster with $U=4t$ and $\mu=0$ in CDMFT (left panel) and DCA (right panel) calculations with 12 bath-sites using different fictitious temperatures $1/\beta$ but constant cutoff $\omega_\mathrm{cut}\approx 200t$ and $N=0$ and 1, starting from different Anderson parameters. For low fictitious temperature results are fairly independent of the starting points, while for larger temperature the Anderson parameters are essentially under-determined by the distance function and consequently the results of the supposedly self-consistent calculation strongly depend on the initial values.}
\end{figure}

To fit the Anderson parameters $V_l$ and $\varepsilon_l$ we use the distance function
%\begin{displaymath}
% \sum_{0<\omega_n<\omega_\mathrm{cut}} \hspace{-3ex}\frac{\left|\left|\mathbf{W}^\dagger\left(\Gbath^{-1}(i\omega_n)-i\omega_n-\mu+\mathbf{H}_c+\sum_l \frac{\mathbf{V}_l\mathbf{V}_l^\dagger}{i\omega_n-\varepsilon_l}\right)\mathbf{W}\right|\right|_I}{\omega_n^N}
%\end{displaymath}
\begin{displaymath}
 \sum_{0<\omega_n<\omega_\mathrm{cut}} 
 \frac{\left|\left|\mathbf{W}^\dagger\left(\Gbath^{-1}(i\omega_n)-\Gbath_\mathrm{And}^{-1}(\{\varepsilon_l,\mathbf{V}_l\}; i\omega_n)\right)\mathbf{W}\right|\right|_I}{\omega_n^N}
\end{displaymath}
where $||\cdot||_I$ is the 1-norm for the block of irreducible representation $I$ and $N$ determines how strongly large Matsubara frequencies $\omega_n$ are weighted. The distance function will only be finite in the limit $\omega_\mathrm{cut}\to \infty$ if the sum-rules are fulfilled exactly, or if $N\ge2$. Thus for $N<2$ increasing the cutoff emphasizes the sum-rules. The same is true for decreasing $\beta$. If the distance function is dominated by the large frequency asymptotics, the optimization mainly focuses on just the sum-rules. 
% role of the tail: sum-rule. distinguish selection rule (sum-rule=0) and sum-rule.
This means that in practice the Anderson parameters can become under-determined. Since the fictitious temperature is only used for fitting, while the calculations are actually for $T=0$ there is a strong dependence of the physical quantities on the Anderson parameters. Hence for small $\beta$ the self-consistent results strongly differ for different initial Anderson parameters. This is illustrated in figure \ref{spread_all}. Interestingly, the situation is opposite to calculation at finite temperature, where at higher temperature the physical quantities become less dependent on the details of the fitting.\cite{aleprivate}

To avoid an under-determination of the Anderson parameters, it is important to ensure that the features of the bath Green matrix close to the real axis are properly weighted in the distance function. This was already pointed out in reference \onlinecite{esperti}. We find that a good compromise between fitting the large- and small-frequency behavior is given by $\beta=256/t$, $N=1$, and $\omega_\mathrm{cut}=200\,t$, which, if not explicitly specified otherwise, is used in the calculations reported in this work.

\subsection{Convergence with number of bath-sites}
\begin{figure}
 \center
 \includegraphics[width=3in]{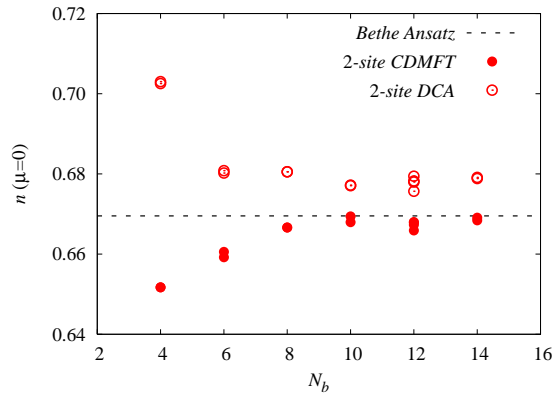}
 \caption{\label{convNB}
  (Color online) Density as a function of the number of bath-sites $N_b$ for a 2-site cluster for $U=4t$ and $\mu=0$. To give a measure of the reproducibility of the results, we plot the densities of several converged CDMFT and DCA runs (multiple symbols at the same $N_b$).}
\end{figure}

To check how many bath-sites we need to reach a satisfactory fit of the bath Green matrix, we consider a 2-site cluster for increasing $N_b$. As example we show in  figure \ref{convNB} the density for $\mu=0$. For CDMFT we find that we need at least 8 bath-sites to obtain a converged density. Also for DCA we obtain convergence for $N_b=8$. Both these results translate to 4 bath-sites per non-vanishing element of the bath Green matrix. It is interesting to note that DCA converges to a density above the Bethe-ansatz result. This could be an artifact of the 2-site cluster, for which the coarse-grained hopping is larger than $t$, because for periodic boundary conditions the hopping on the cluster and across the boundary add up. Averaging over different choices of boundary conditions or going to a larger cluster might improve the situation.\cite{jarrellprivate}

\begin{figure}
 \center
 \includegraphics[width=3in]{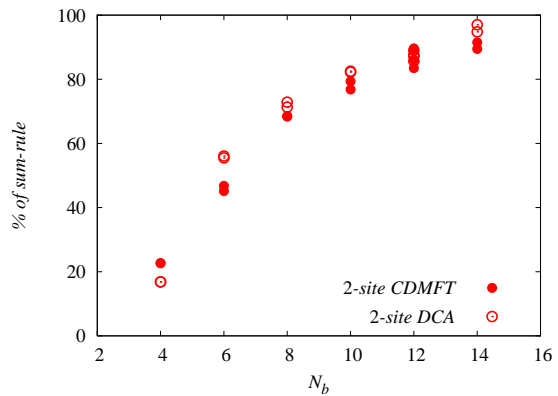}
 \caption{\label{convsumr}
  (Color online) Fraction of hybridization sum-rule for a 2-site cluster as a function of $N_b$. As more and more bath-sites are added, the exact hybridization sum-rule is approached. Shown is the percentage of the sum-rule for irreducible representation $A$ in CDMFT and $K=0$ in DCA. For irreducible representation $B$ and $K=\pi$ we find very similar behavior.}
\end{figure}

Checking the hybridization sum-rule for the diagonal elements of the bath Green matrix, we find that the density is already converged while the sum of the hybridizations $\sum_l |V_l|^2$ is only at about 70\% of its exact value. As shown in figure \ref{convsumr} we need to go to even larger baths to properly fulfill the sum-rule. We tried also to fit the bath Green matrix imposing the sum-rule, i.e.\ fixing one Anderson parameter per irreducible representation. We found, however, that for small $N_b$ this does not give particularly good results, as it weights the large frequency behavior of the bath Green matrix too strongly, while for large $N_b$ it is not necessary.
%The off-diagonal elements of the bath Green matrix vanish in our DCA parametrization by construction. In the CDMFT parametrization with irreducible representations we find that already for small $N_b$ the bath sites of $A$ and $B$ symmetry are essentially degenerate, implying that the off diagonal elements of the bath Green matrix vanish.

Going to larger clusters, we expect that we will need more bath-sites for a converged calculation. A notable exception is CDMFT for the linear chain with nearest-neighbor hopping only. As discussed above, in this case there are two identical baths coupling to one surface site each. Since the bath Green function that these baths have to fit is fairly independent of cluster size (see figure \ref{CDMFTbath}) we expect that the number of bath-sites needed for convergence is independent of cluster size.

\subsection{Convergence with number of cluster-sites}
\begin{figure}
 \center
 \includegraphics[width=3in]{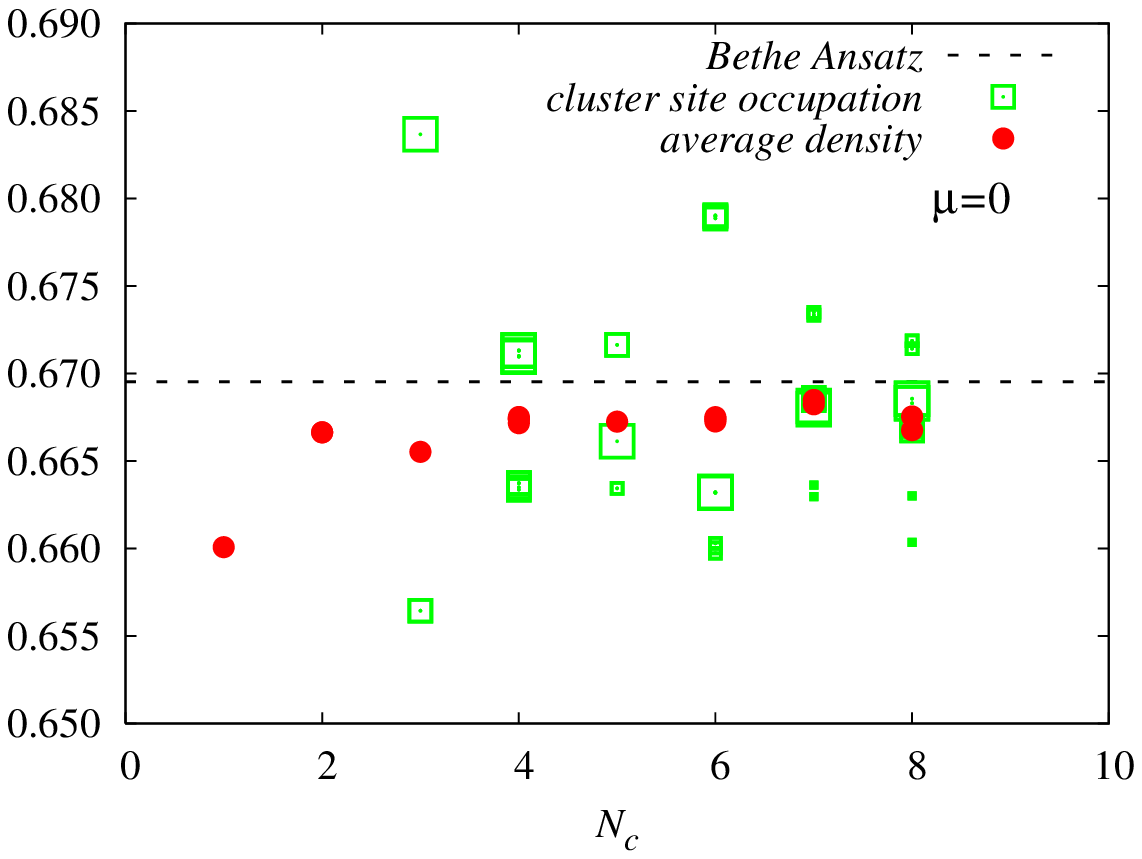}
 \includegraphics[width=3in]{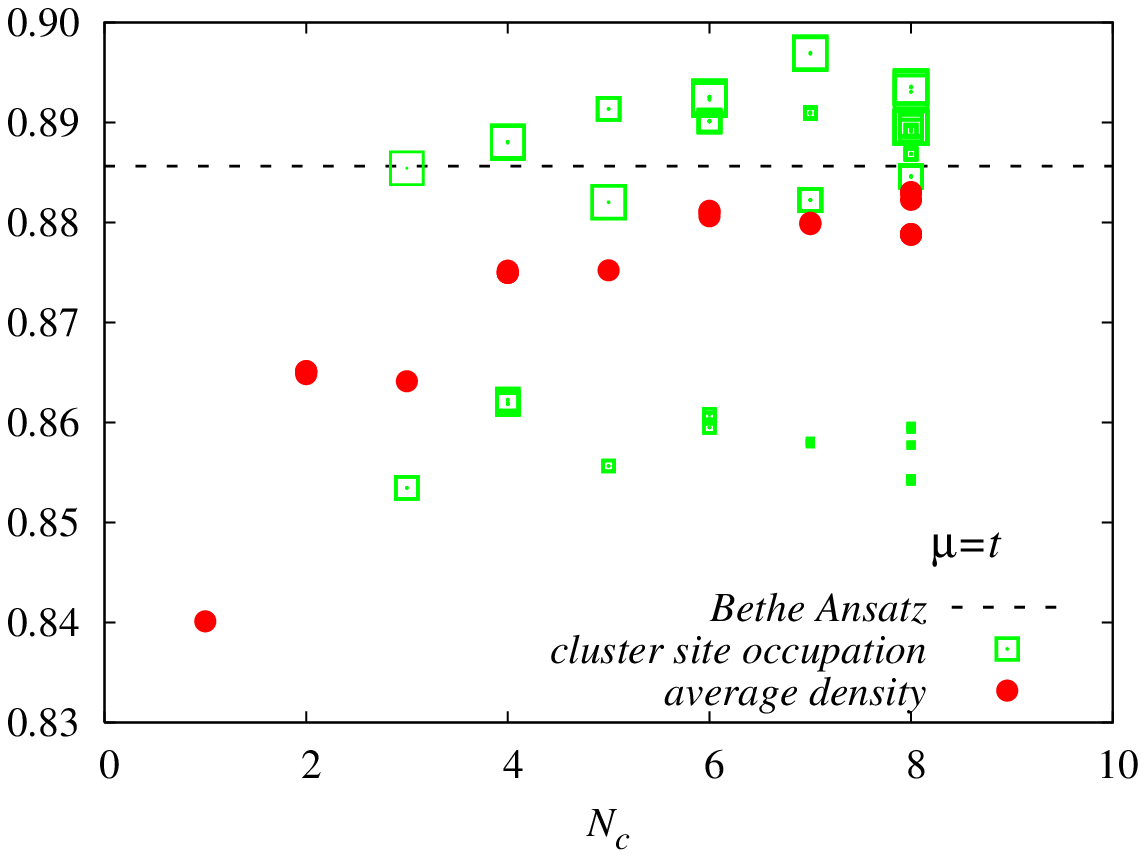}
 \caption{\label{convNC}
  (Color online) Density for linear CDMFT clusters of increasing size $N_c$ with $U=4t$ at $\mu=0$ (upper panel) and $\mu=t$ (lower panel). All cluster calculations are for $N_b=8$. Circles denote the average density per cluster-site. (Green) Open squares are the individual cluster-site occupations. The size of the squares indicates how close the site is to the center of the cluster. To assess the reproducibility of the calculation we show for each even value of $N_c$ the result of at least two to seven runs with different choice of the starting set of Anderson parameters. The dotted line represents the exact Bethe-ansatz result.}
\end{figure}

We now analyze the convergence of CDMFT with the number of cluster sites $N_c$. As before we focus on the density $n$, which is shown in figure \ref{convNC} for $U=4\,t$ and chemical potential $\mu=0$ and $\mu=t$. Considering the series of odd or even $N_c$ separately, we see that with increasing cluster size the average density systematically approaches the exact result for the infinite chain. Interestingly, going from a cluster with an even number of sites $N_c$ to $N_c+1$ the average density hardly changes.
For both chemical potentials already the smallest cluster gives a significant improvement over a single-site calculation. For $\mu=0$ the exact density is basically obtained for $N_c=2$. For $\mu=t$ convergence to the infinite-chain result is only reached at $N_c=6$. 
% small odd clusters -- frustrated RVB???
    
\begin{figure}
 \center
 \includegraphics[width=3in]{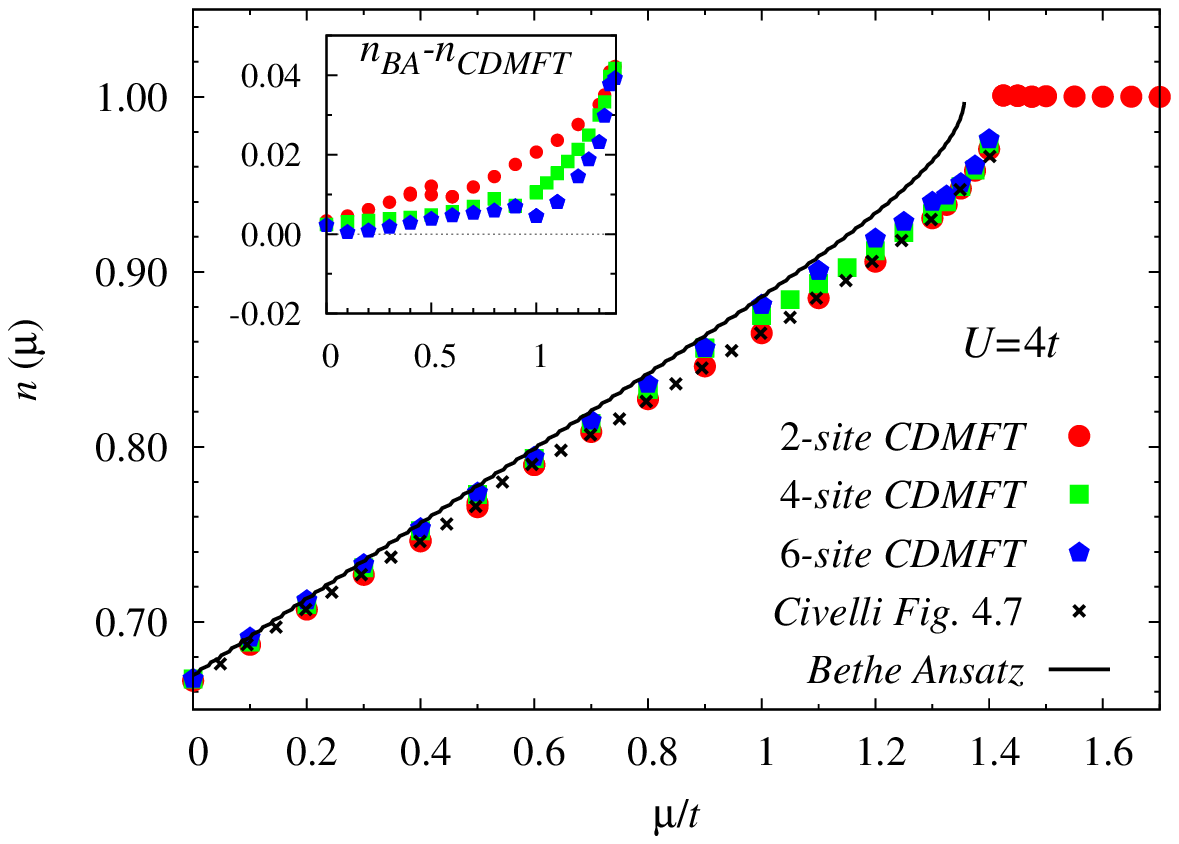}\\
  \includegraphics[width=3in]{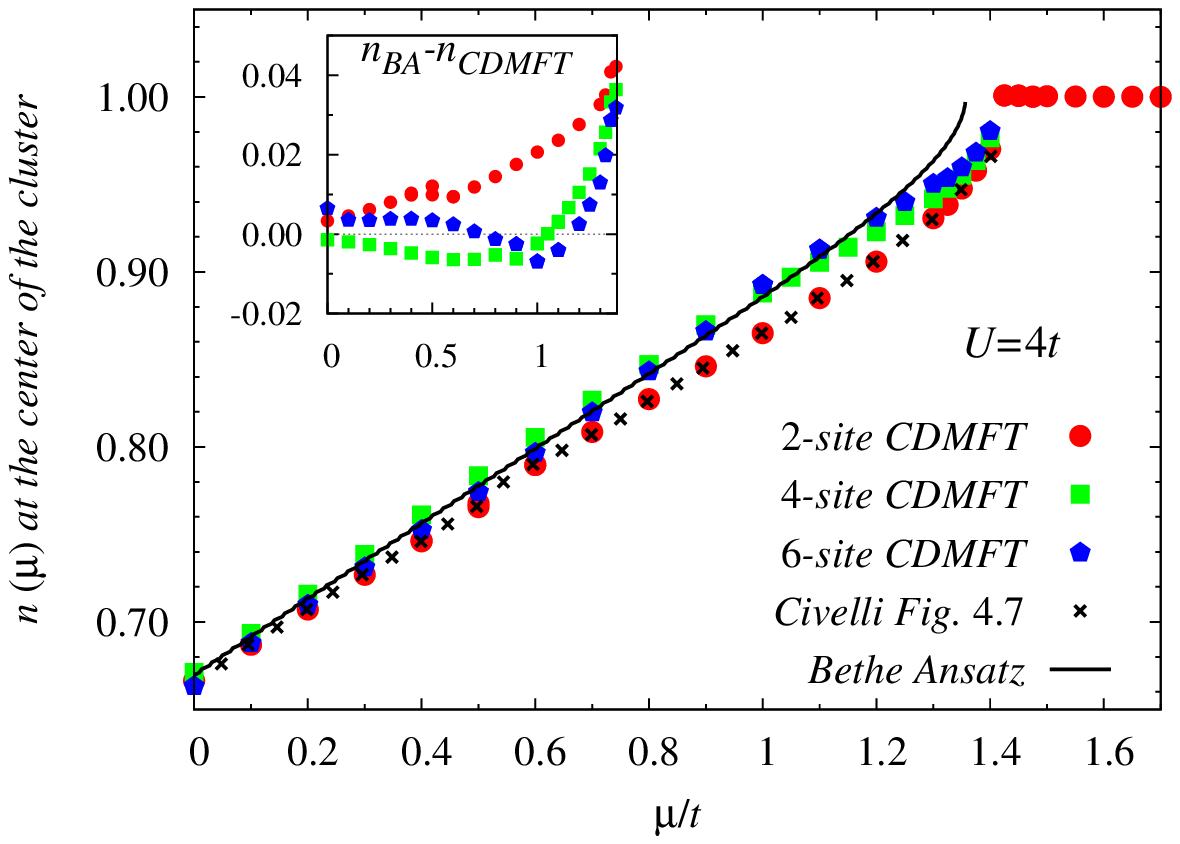}
 \caption{\label{esperto}
  (Color online) Density $n$ as a function of chemical potential $\mu$ at $U=4t$ for linear CDMFT clusters of increasing size and $N_b=8$ compared to the exact result for an infinite Hubbard chain (Bethe Ansatz). The crosses give the results from figure 4.7 of reference \onlinecite{civelliPhD}. The inset shows how the convergence of the density with increasing cluster size to the Bethe result becomes progressively slower close to the Mott transition. $n_{BA}-n_{CDMFT}$ jumps when the self-consistent solution changes sector $(N_\uparrow,N_\downarrow)$. Close to these sector changes the results slightly depend on the initial conditions, i.e., there is a hysteresis between calculations increasing or decreasing $\mu$. This is shown as multiple symbols for a given chemical potential. The upper plot shows the average density for 2-, 4-, and 6-site clusters, the lower plot the density on the two central sites.}
\end{figure}

In figure \ref{esperto} we show how the density versus chemical potential curve for CDMFT calculations of increasing cluster size approach the exact result for the infinite Hubbard chain. We find that the closer we come to the metal-insulator-transition the harder it gets to reach the infinite-size limit. This does not come as a complete surprise, as the self-energy is expected to become strongly $k$-dependent at the Mott transition.\cite{mottself,Stanescu} 
% sector changes:
% 2-site: mxu=0.5 (4,4) --> xmu=0.6 (5,5)
% slight hysteresis:
% mu    n          g.s. sector
% upwards:
% 0.4  .7464880207      4,4
% 0.5  .7655889995      4,4
% 0.6  .7896430557      5,5
%
% downwards:
% 0.6  .7896429451       5,5
% 0.5  .7677706434       5,4 
% 0.4  .7460220110       4,4
%
% 4-site: xmu=0.8 (5,5) --> xmu=0.9 (6,6)
% 6-site: xmu=0.9 (6,6) --> xmu=1.0 (7,7)

\begin{figure}
 \center
 \includegraphics[width=3in]{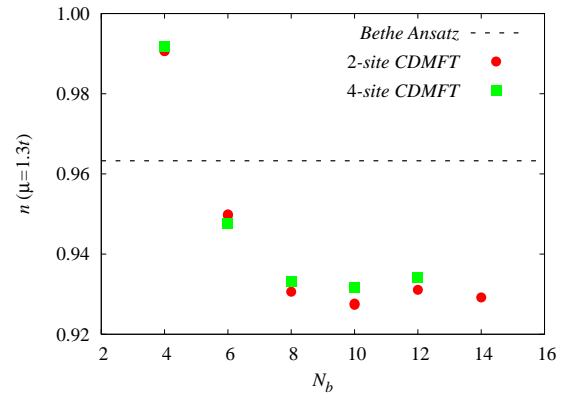}
 \caption{\label{NB13}
  (Color online) Average density as a function of the number of bath-sites $N_b$ for 2- and 4-site clusters with $U=4t$, $\mu=1.3\,t$, and $\beta=512/t$. As for smaller chemical potential the density is converged for $N_b=8$ (cf.~figure \ref{convNB}). We observe that for larger $\mu$ unconverged calculations with small baths, e.g.~$N_b=6$, can give densities closer to the one of the Hubbard chain.}
\end{figure}

We note that our results for $N_c=2$ agree with figure 4.7 of Ref.~\onlinecite{civelliPhD}. They are {\it not} compatible with figure 4.1 of Ref.~\onlinecite{civelliPhD} and figure 2 of Ref.~\onlinecite{esperti}. We have checked that our calculation is properly converged by starting from a number of different initial points, always converging to essentially the same density. To achieve this, for chemical potentials $\mu\ge1.3\,t$ we increased $\beta$ from $256/t$ to $512/t$, in line with the trend shown in figure \ref{spread_all}. We can, however, get significant variations in the density by putting restrictions on the bath parametrization. Using, e.g., only six bath-sites, the CDMFT result happens to be closer to the one for the infinite system, in the vicinity of the Mott transition.\cite{civelliPhD} 
This is shown in figure \ref{NB13}. For small baths the calculated density is very sensitive on $N_b$ and can be either larger or smaller than the density for the infinite chain. For such small baths results will therefore critically depend on the fitting. 
We can, e.g., artificially ``improve'' the result by forcing a pair of bath energies to zero. Other restrictions on the bath-parameters instead move the densities further away from the Bethe curve. In all these cases we find that the restricted bath-parametrization results in a significantly deteriorated fit of the bath Green matrix. I.e., the sensitivity of these calculations to technical details merely shows the effects of an inadequate fitting of the bath Green matrix.
By increasing $N_b$, the bath-parametrization improves and the calculated density converges, as shown in figure \ref{NB13}. Nevertheless, differently from the behavior at smaller chemical potential (cf.~Fig.~\ref{convNB}), the converged value is substantially smaller than the Bethe ansatz one. As this does not improve much with increasing $N_c$, we can conclude that the clusters are still too small to accurately capture the behavior of the infinite system close to the Mott transition.
% This is also supported by 2-site CDMFT calculation using a different solver (QMC at finite temperature)\cite{kyungQMC} and by variational cluster approximation (VCA).\cite{VCA}

\section{Conclusions}\label{SecConclusions}
The central problem of dynamical mean-field calculations with a solver that uses a small number of sites is to find a good parametrization of the bath. To address this problem for dynamical cluster approximations we have presented a systematic formulation for the bath degrees of freedom. We have found sum-rules which allow to identify what hybridizations vanish and hence need not be parametrized at all. In addition the sum-rules can be used to check convergence for small baths. For the non-vanishing hybridization functions, we have introduced a bath-parametrization based on the irreducible representations of the cluster Green matrix. In this approach the fitting of the bath sites is broken into independent fits of irreducible blocks of the Green matrix. This leads to a significant simplification of the fitting procedure which is particularly important when dealing with large baths. The symmetry-based approach should also benefit the variational cluster approximation (VCA),\cite{VCA} where the determination of the parameters requires a Lanczos calculation in each optimization step. 

As an application we have revisited the Hubbard chain. While this one-dimensional problem is the worst case scenario for DMFT, which is exact in infinite dimensions, it is technically the easiest case for CDMFT, because it requires only a minimal bath which is essentially independent of cluster size. This allowed us to study the results of CDMFT using linear clusters of increasing size, extending previous work that was limited to 2- and 3-site clusters.\cite{esperti,civelliPhD} Analyzing the density as a function of chemical potential, we find that results significantly improve already going from a single-site DMFT to a 2-site cluster and become systematically better for larger and larger cluster sizes. Close to the Mott transition the convergence with $N_c$ critically slows down, implying that the $k$-dependence of the self-energy gets more and more important. 

\begin{figure}
 \center
 \includegraphics[width=3.3in]{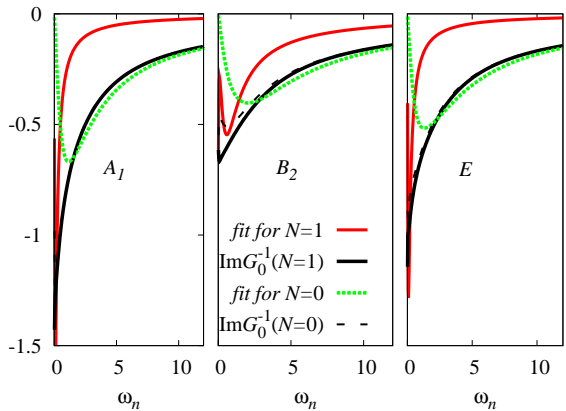}
 \caption{\label{2by2bath}
 %n=0.525 (N=1), 0.562 (N=0)
 (Color online) Fitting the bath Green function for a 2$\times$2 cluster in CDMFT with $U=12\,t$, $\mu=0$, $N_b=8$. The imaginary part of the block-diagonalized bath Green matrix (shown are only the three inequivalent blocks labeled by the irreducible representation) and its corresponding fit are plotted for two values of the exponent $N$ (cf.~section \ref{fitting}). The quality of the fit is clearly very poor in both cases (for $N$=1 compare full red/gray line with full black one and for $N$=0 compare dashed green/gray line with dashed black one).}
\end{figure}
The fortuitous independence of the CDMFT bath on the cluster size for the Hubbard chain is lost in higher dimensions. Already for a cluster as small as $2\times2$ we have to fit three functions, one of which belongs to a doubly degenerate $E$ representation. In that case $N_b=8$ translates to only two effective bath-sites per bath Green function. For comparison, in our one-dimensional calculations we need at least 4 bath sites per bath Green function for a converged bath. Hence, using a Lanczos solver, it is virtually impossible to converge a $2\times2$ CDMFT calculation. This is illustrated in figure \ref{2by2bath}.

The same is true for DCA, since the number of baths increases with cluster size, independently of the dimensionality of the cluster or the nature of the hopping. For zero-temperature cluster calculations it is therefore mandatory to move to impurity solvers that can handle large baths, e.g., DMRG.\cite{DMRGsolver} For these calculations with large baths the efficient parametrization of bath and fitting of irreducible blocks will become even more important.

\section*{Acknowledgments} 
We would like to thank M.~Capone for inspiring the present systematic treatment of CDMFT and for important discussions in the early stage of the work. M.~Civelli kindly shared with us many details from his PhD thesis. A.~Parola and S.~Sorella provided us with the density curve for the Bethe Ansatz solution. We also acknowledge useful discussions with C.~Castellani, M.~Hettler, M.~Jarrell, S.~Kancharla, G.~Kotliar, A.I.~Lichtenstein, I.~Mazin, M.~Potthoff, S.~Sakai, and A.~Toschi. S.G.~thanks the Forschungszentrum J\"ulich for hospitality. Calculations were performed in J\"ulich on the JUMP computer under grant JIFF22.


\begin{thebibliography}{99}
\bibitem{accdeg} 
 In case of accidental degeneracies, we can always symmetrize the degenerate 
eigenstates.
\end{thebibliography}
\end{document}